# Data Integrity and Dynamic Storage Way in Cloud Computing

Dinesh.C, *P.G Scholar, Computer Science and Engineering, Mailam Engineering College*, *Mailam, Tamilnadu.*

*Abstract*—It is not an easy task to securely maintain all essential data where it has the need in many applications for clients in cloud. To maintain our data in cloud, it may not be fully trustworthy because client doesn't have copy of all stored data. But any authors don't tell us data integrity through its user and CSP level by comparison before and after the data update in cloud. So we have to establish new proposed system for this using our data reading protocol algorithm to check the integrity of data before and after the data insertion in cloud. Here the security of data before and after is checked by client with the help of CSP using our "effective automatic data reading protocol from user as well as cloud level into the cloud" with truthfulness. Also we have proposed the multi-server data comparison algorithm with the calculation of overall data in each update before its outsourced level for server restore access point for future data recovery from cloud data server. Our proposed scheme efficiently checks integrity in efficient manner so that data integrity as well as security can be maintained in all cases by considering drawbacks of existing methods

## I. INTRODUCTION

Cloud computing is a virtualized resource where we want to store all our data with security measurement so that some application and software can get full benefits using this technology without any local hard disk and server for our data storage. These services are broadly divided into three categories as

1) Infrastructure-as-a-Service. 2) Platform-as-a-Service and 3) Software-as-a-Service [1] [7]. It is not so efficient by using a few technologies such as flexible distributed scheme, Reed Solomon technique, and BLS algorithm [8] to give more integrity to the operation like change, removal, and add of data from cloud server. So to offer yet sophisticated and efficient integrity to cloud data so that all applications, software and highly valuable things can't be affected and modified according to unknown person's challenge from un-trusted cloud we have to provide more integrity method apart from previous methods with the assist of CSP who maintains our data from cloud servers from their IP address domain [3]. But since we may not have copy of outsourced data CSP can behave falsely regarding the status of our outsourced data. And since we don't have any physical data control we have to fully depend on CSP for cloud data access without any separate login for our self access in data management. So we have some limitation in this view so that CSP only can take more rights to manage our data with his IP address domain. But here we need to keep up data integrity for our cloud data in efficient manner for our own usage whenever there is a want for that and must have some integrity measurement for our data storage to protect data from internal and external attack including byzantine as well as malicious attack. And in this case, there are some different dealings such as file distribution algorithm, error location correctness and data authentication [8] using distributed protocol for its integrity as far as existing system is concerned. But all these appear from existing system with a few restrictions without learning data integrity in competent manner using effective automatic data reading protocol from user as well as cloud level before and after the data append into the cloud. And also we have done multi-server data comparison algorithm for every data upload for making efficient way for data integrity. If there is any server failure, then the data can be recovered repeatedly in cloud server using this scheme.

## II. ACTIVITIES OF THE CLOUD

### A. Devoted Property Assertion

All resources have high-quality assurance in its multi cloud servers with the help of strong platform. The mentioned possessions of RAM, CPU, and Bandwidth capacity [4][5] in network produce sufficient atmospheres for its concerned usage. Thus the activity of could server can't ever opposite to the current world while considering its level of integrity and security even though there is some little bit of conflict that we want yet to recover from its already existing level.

### B. Excess Data Storage

All cloud server storage resources are managed by high-performance and high-availability storage area network. Many cloud solutions operate on local disks from the host system, which means any computing or storage failure can result in down time and potential data loss. As cloud servers are autonomous, if there happens any server crack in stored data, these can be protected against internal and external attacks.

### C. Client Support and Stability

It clearly tells us that user can do the increase and decrease of the data capacity in the cloud server with the help of CSP (cloud service provider) in his request. This storage level must be with flexible and durability condition as far as its entire design or structure is concerned. Thus it should be claimed extra storage space concerning future process in data exchange.

### D. Pay Costs for Access

Customers have to be charged according to their usage and their data storage in cloud storage apart from allocated space. Customers can charge depending on the usage of network access for data exchange with good bandwidth [4] [5] as well as data storage. For example, it is just like when utility company sells power to consumers and telephonic company offering utilities for their services.

### E. Efficient Dynamic Computing Environment

Data should be in a self-motivated computing infrastructure. The main concept in this dynamic environment is that all standardized and scalable infrastructure should have dynamic operation such as modification, append, and delete. The cloud platform which has virtualized conditions also should have some specific independent environment.

### F. Accurate Choice of CSP

To get excellent service from multiple servers, good service providers are important to be considered and selected. So much care must be taken in this respect so that the CSP itself can be elastic with clients in order to get accessed with all places (anywhere and anytime). It has the following benefits such as,

  1) Cost savings- To save expenditure among IT capabilities.
  2) Trustworthiness- Data back up by CSP in cloud servers if system is stolen and loses the data itself.
  3) Scalability on requirement- Whenever there is a need for data accessing, user can be easily accessible to that anytime and anywhere.
  4) Protection expertise- Cloud service providers in general have more skill securing networks than individuals and IT personnel whose networks are usually associated to the public Internet anyway.
  5) All over Access- It allows users ubiquitous access to computing power, storage, and applications.

### III. PROBLEM REPORT AND SYSTEM STRUCTURE

This has following system formation for its structure as 1) User 2) Multiple cloud servers 3) Cloud service provider 4) Data reading protocol 5) Server access point.

### A. Making Server Access Point and Time Intimation

To keep the data away from server failure in every data inclusion by unauthorized person or any internal and external attack coming within CSP address domain, one access point or restore point in every update is given to the cloud server when client does some delete, modification, and append in his will. It is done in the time of data comparison in every update for the data by user with the help of CSP. This is a new technique we have introduced in our system design. The reason for which we want to do this restore point is that if there is any server crash then this restore access point helps a lot to recover the whole thing that we have lost. This process is proceeded by fixing one access point to cloud server database in update or data upload (it is done with the help of CSP, because we don't have any replica of our outsourced data) when we finally terminate our data exchange in cloud server. Also since we don't have any physical custody of our data in cloud server we can't have any separate login access using cryptographic key. It is a major disadvantage to our cloud server to maintain our individual or group data. In coming days it can be improved with the help of CSP. When user uploads the data into the cloud automatically these data are included and created for access point and here one specific restore point is done in its cloud server for effective recovery purpose.

### B. Optional Third Party Auditor

To reduce the work load of the user, (see Figure 1) TPA [1] [8] has the delegation for the data checking with a few limitation so that he can't modify content of the data in his auditing. And TPA pays due care on storage correctness verification.TPA auditing is in the manner of privacy preserving concept so that verification can be done in separate manner alone from any interaction by others. In the existing system, distributing protocol is used for this purpose. Here TPA does not have special authority for data integrity checking. However we don't give more importance to TPA to such a great degree but a small bit only since our proposed system pays more concentration on client's access for full security.

### C. What the Existing System Has?

Many expertise's express us that whenever opponent tries for any attempt to append, delete and modification for cloud data from its storage level it is protected from that (byzantine problem, malicious, data modification attack, and also server failure) [8] using flexible distributed storage integrity auditing mechanism with distributed protocol. And it tells us that whatever changes has occurred by the above stated failure it is cleared only using specified file distribution and error correctness method. Typically it concentrates mostly on TPA auditing other than the client. Token format is followed for data arrangement in array format for their task to be performed. All Security measurement is processed through state-of-the-art.

### D. Disadvantages of Existing System

Even though it concentrates more on time consuming for user almost user has to depend on TPA for data integrity whenever he doesn't have time for auditing and he automatically can't have the any stored data size after any such malicious, byzantine and system failures to know about data modification, delete and append in his own knowledge. But it does not tell about its (data) fixed storage capacity for user's stored data in cloud before and after in cloud storage area or in server (how much of data has been stored in server for particular clients in his own allotted storage server area). So it is a major issue to user and also almost users have to depend on CSP for extensive security analysis and depend partially on TPA. It doesn't tell effective way for server failure. It doesn't make efficient rout for data integrity whenever user want required data from his earliest storage in cloud server. But here complete access is going to CSP. So CSP can behave in its own way by hiding any loss of data since existing system doesn't tell about the weight or value of stored data using any algorithm.

### E. Our Proposed Model

We suppose that CSP allotted space is a major concern for our data maintenance in all manner for dynamic operations. All outsourced data and data entering into the cloud is measured using reading protocol algorithm. So in order to keep integrity of overall data, we have to use "data reading protocol from user as well as cloud storage level before and after the data adding into the cloud server area' and another "multi-server data comparison algorithm for every data upload for the purpose of data recovery management" for our proposed achievement. When server failure occurs in cloud entire data may be affected so that user can't foresee data's trustworthiness in its whole atmosphere depending on variety of situation or CSP's process to hide the loss of data. Taking into account these, we have modeled and proposed automatic protocol and server data management algorithm to know about entire data exchange before and after the data insertion into multiple cloud server. And user can know if there has been done any change, remove, and attach operations that can occur for data from its storage level with the help of our proposed scheme that never has been told. It can be effectively administered by clients from appropriate efficient data reading protocol from cloud server position.

### F. Advantages of Proposed Model

By processing the integrity of data using data reading protocol and data management algorithm after and before the entering of data into the cloud, user can assure that all data in cloud must be in protected condition for its trustworthiness. So easily the actual size of stored data before and after in cloud is maintained even though the user himself has done any modification, deletion, and update for his own purpose by using proposed scheme. These processes are carefully done using our proposed scheme. So here user takes full control and process on the data stored in cloud apart from TPA and we can give strong assurance and protection to the data stored in multiple cloud server environment. To avoid server failure and any unexpected error we should put one server restore point in cloud server database for efficient data back up or restore using multi server data comparison method. It is major advantage of our proposed system. This process is done with the help of CSP for cloud database process since we have physical data possession in cloud server.

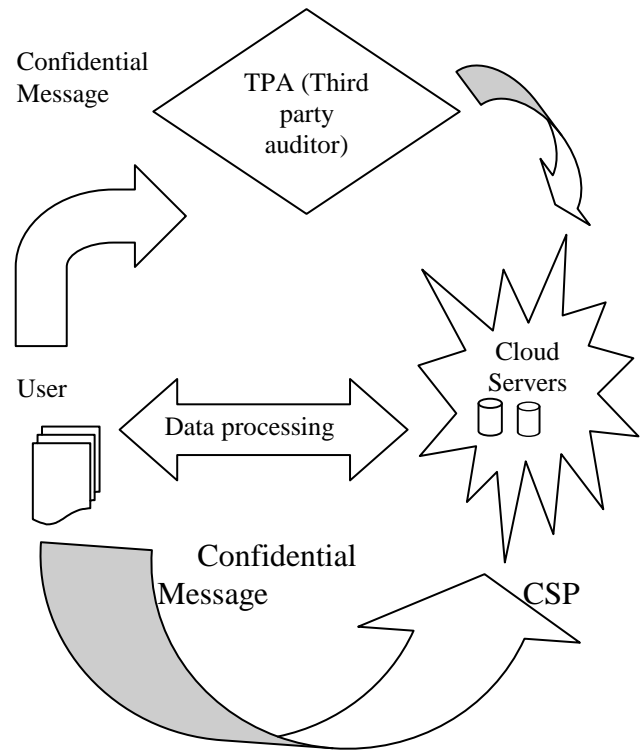

Figure 1: Cloud Architecture and Data Processing Way.

## IV. PROCEDURE FOR DATA VERIFICATION

Assume that the data to be uploaded is 50GB, or let it be 'N' GB and N is divided into as many parts. if it is divided into (n-1) of servers, then $n \in N$ and $n_i, n_i+1, (i-1) \in N$ and All these data are stored in G, then G={ $n_i, n(i+1), \ldots n(i-1)$} then G=>N, where n is number of node or server for data. Here one signature is provided for this procedure to conform the fixed integrity of the data in its cloud server. And client and server have separate input for its further action. Here we have number of server for the data upload operation in its input level. It is received from client running position. To maintain integrity of data we should compare the data from its user level and its reach of cloud.

### A. Data Verifiability in User Level before Upload

In initial stage upload the data, this process expected from its starting condition for "N" and "G" These are number of server and cloud storage fixed server respectively. The verification of data is done here from user level before sending or uploading data to destination or cloud server. it is corresponding to the value of the same data after its reaching of cloud server or in its cloud level. Then it is verified using data read protocol in both server and user level. If the values are not changing from its position it is confirmed that all data have been maintained in well secured manner. So in order to maintain integrity of data as already we told here, before sending data into the cloud server it is verified for its value in user control. Then, G= {$n_i,...n_{i+1},....n_{(i-m)}$} and The following equations explain it in clear way as,

1) n ∈ G, ∀n, here n>1>0 where G is cloud level identification and n is number of server from user level.

2) n={$i_a,i_{a+1,.....}i_{a-n}$}, where a is data counting for n server.

3) n={{$R_1${ $i_a,i_{a+1,.....}i_{a-n}$},$R_{1+1}${$i_a,i_{a+1,.....}i_{a-n}$},…$R_{i-n}${ $i_a,i_{a+1,.....}i_{a-n}$}}} , where R is the number of different server divided into the cloud.

4) ∀n, n= ($R_{1i},..R_{2i},…R_{n-i}$), it is up to the 'R' number of server count

5) ∀i, i=any data in G and while checking for integrity it gets "n=>R" formation. It means that n implies R in all manners.

### B. Algorithm in User Level

1: //Let "G" be the storage or cloud
2: //"N" the total number of server
3: //Input n and I for server and counter
4: //G={$n_i,n_{i+1},……..,(n_{i-n}$-1)}
5: //if(n<=G)&&(n==G)
6: {
7: for(i=0;i<=n;i++)
8: S=i;
9: S++;
10: }
11: else if
12: (n==R)
13: {
14: or (i=0;i<=n;i++)
15: {
16: n++;
17: }
18: stop the program.

### C. Checking for Same Data in User and Cloud Level

Initially we should assume for Boolean value for true and false condition by comparing both user and cloud level data. Then,

1) Verify ($Z_i$={$G_{1i}$(true)$G_{2i}$(false)})

Check for condition,

2) if the condition $G_{1i}$ = $G_{2i}$ then it is true in such a way that,

3) n==$n_u$ it is set that Z={s(true)}

n!=$n_u$ otherwise,

4) Z=={s(false)}. These are the conditions for maintaining the user and cloud level data integrity for all user and server data, since the case has the above condition every user and cloud lever data comparison has different.

### D. Data Verifiability in Cloud Server Level after Upload

Here as we did before in our user level part the same procedure is followed. The only difference is that all processes are done in cloud server location using verification protocol or data reading protocol. The changes can be as in following manner. Assume that $G^u$=>G. It is done for same data identification among user and server level data content. Then $G^u$ uploaded data into the cloud server.

1) $n^u$∈$G^u$, ∀$n^u$, here $n^u$>1>0 where $G^u$ is cloud level identification and n is number of server from user level.

2) $n^u$={$i_a,i_{a+1,.....}i_{a-n}$}, where a is data counting for n server.

3) $n^u$={{$R_i${ $i_a,i_{a+1,.....}i_{a-n}$},$R_{1+1}${ $i_a,i_{a+1,.....}i_{a-n}$},…$R_{i-n}${ $i_a,i_{a+1,.....}i_{a-n}$}} , where R is the number of different server divided into the cloud.

4) ∀$n^u$, $n^u$=($R_{1i},R_{2i},…R_{n-i}$), it is up to the 'R' number of server count.

5) ∀$i^u$, i=any data in $G^u$ and while checking for integrity it gets "$n^u$=>R" formation. It means that $n^u$ implies R in all manners.

### E. Algorithm in Cloud Server Level

1://Let "G" be the storage or cloud
2://"N" the total number of server
3://Input n and I for server and counter
4://G={$n_i,n_{i+1},……..,(n_{i-n}$-1)}
5://if(n<=G)&&(n==G)
6:{
7:for(i=0;i<=n;i++)
8:S=i;
9:S++;
10:}
11:else if
12:(n==R)
13:{
14:for (i=0;i<=n;i++)
15:{
16:n++;
17:}

### F. Putting Server Restore Access Point for Data Recovery

When server failure happens all data may be lost its integrity and since user doesn't have the local copy of data there is no more possibility to recover the already lost data from its previous original condition. So here we have modeled one scheme "multi-server data comparison algorithm for every data upload for the purpose of data recovery access point in every data update by user". This prevents entire system collapse from data lose against any such type of system crash including Byzantine, and related internal and external problem. The above scheme explains the server crash breaking condition in efficient manner by putting one restore access point in previously updated data from user.

1) Assume server access point for already stored data, X and after the server failure

2) Put automatic restore point then,

3) Compare for previous value with current total value. Now the previous value is as, => $\sum_{i=0}^{n} [(S^d+T^d)]^{d-1}$=X

4) Current value before crash is as, =>$\sum_{i=0}^{n}[(S^d+S^d_1)+(T^d+T^d_1)]$=Y

5) Now when we do comparison for both X and Y, the following assumptions are made as, =>X>Y or X<Y then do update otherwise if X==Y then restore to the same condition. Here this formula is applied for overall data from the server by receiving user and server level data. Here "S" is the cloud level data and "T" is the user level data and "d" is the number of data depending upon the upload condition.

## V. OUR PROPOSED DATA BLOCK DYNAMIC PROCESS

To ensure the assurance of the data we can do the operations such as append, deletion, and update. These are the dynamic data operations to be done in the cloud area by user.

### A. Append Operation in Server Block

We may assume that there is any size of GB space allotted by CSP user's requirement for any application purpose. Then first, this size is calculated and compared using our technique. It is clearly mentioned in our algorithm specification. In the comparison the storage cloud area is confirmed that it does not have any data in its position for strong integrity. Also the operations such as add, change, and removals by client are processed by space measurement scheme for effective identification of data integrity in cloud database. So if there is any such modification by attack then client can give assurance to the data integrity by successfully following. It is very efficient method in our proposed design compared to any other such type.

### B. Deletion Operation in Server Block

First, we want to compare value from existing cloud server. Then, this deletion operation depends on user's attempt on his data stored in cloud server using his login operation. The following operation is performed in this deletion operation. If there is number of servers for data selection while deleting the data stored the particular storage server is considered for this. The required data can be deleted using above mentioned algorithm.

### C. Update Operation in Server Block

This operation is finally finished after completion of needed action by user. And for update, the above derived algorithms are taken into the account. For new data update, each data block should be updated automatically from already existing value to new updated value. So if we consider this data block as one array formation then, the result is as $S = (S_i \pm D_i)$ depending on the user's operation on data. It may be positive or negative depending on the update operation to be performed.

## VI. PUBLIC VARIABILITY AND TRUSTWORTHINESS

While considering our paper, we have not considered TPA greatly for data integrity purpose since this task almost is completed by user himself. Only the limited authority is given the TPA so that he can't be able to modify the contents of user's data in strong protection. So he can't operate on his own desire for outsourced and in-sourced data in cloud.

## VII. SOME OTHER TYPES OF ATTACKS THROUGH INTERNET LAYER

This attack is based on the web server and network configuration in relevant attacks. For this attack, we can use some method of reverse sweep by coming to the same source from destination for its correctness. Here protocols such as HTTP, SOAP and FTP protocols [2] are considered for clearance. Also this attack is based on the router. Normally the data packet transformation is prevented by these attacks.

## VIII. BENEFITS OF DATA ESTIMATION FROM CLOUD SERVER

While considering all other cloud concept from different authors, here we have proposed new scheme, "storage management algorithm with time management" for the data calculation to maintain integrity of data. It is one efficient and different method to protect from major data modification, deletion, and append. The main purpose of our proposed scheme is that the total capacity of our allocated space after and before the date update in cloud server is accurately measured by our defined algorithm for the purpose of checking how much of data we already had before any server failure or any internal and external attack. We can maintain integrity of data in well secure manner than other defined methods where the existing systems have.

## IX. CONCLUSION

From our study we have good conclusion of the integrity and security of cloud level data when user do some update such as attach, removal, and change at own desire. So here user mostly has the situation to believe the service provider in many manners. Understanding all these situations, we have handled new proposed technique for cloud storage and. This is the modern way we have used in our paper different algorithm compared to previous related paper in this respect for cloud management. From this protection for cloud data, user can be strong belief for his uploaded data for any future purpose or his any other related process without worry. Here complicated, internal, external and malevolent attack is known by our proposed scheme in efficient manner by storage data measurement (i.e., since there can be some modifications in cloud data). Thus our main idea is to give integrity to the cloud storage area with strong trustworthiness so that user can feel free of worry for his uploaded data in his allocated space. Here our scheme ensures for any extra inclusion of unwanted bits or related things in cloud area so that they can be so easily found out by our data measurement concepts in efficient manner. And it finds out how much of changes have occurred there in its cloud area.

## X. FUTURE ENHANCEMENT

Here we leave more ways as Future enhancement to process for maintaining security and integrity of data using read and write protocol for data calculation from cloud data storage in days to come so that user can identify inserting the attempt of different data having same weight in un-trusted cloud server. This read and protect protocol is efficient to identify any data modification into the server with accurate reading and protecting capacity automatically when such attempt is made by known one. In our future study we also have planned to implement the locking protocol in cloud data storage with the help of CSP for data update. It can give clear security to user's own data when users complete his requirement.

**Dinesh.C** is a post graduate student pursuing Computer Science and Engineering from MAILAM ENGINEERING COLLEGE under ANNA UNIVERSITY OF TECHNOLOGY. His area of interest is security and privacy in Cloud Computing, and Secure Mobile Cloud as well as Cryptography and Network security. He is currently doing his project in cloud computing area, and has published **one international journal, International Journal of Computer Applications (0975 – 8887) Volume 31– No.6, October 2011, in Cloud Computing**.